\documentclass[prd,aps, tightenlines, preprintnumbers, showpacs, nofootinbib,superscriptaddress,notitlepage,11pt]{revtex4-1}

\usepackage{graphicx}
\usepackage{amsmath,amsthm,amssymb}
\usepackage{mathtools}
\usepackage{siunitx}

\newcommand{\beq}{\begin{eqnarray}}
\newcommand{\eeq}{\end{eqnarray}}

\begin{document}

\preprint{YITP-17-02}

\title{
Elliptic Flow in Small Systems due to Elliptic Gluon Distributions?}

\author{Yoshikazu Hagiwara} 
\affiliation{Department of Physics, Kyoto University, Kyoto 606-8502, Japan}

\author{Yoshitaka Hatta}
\affiliation{Yukawa Institute for Theoretical Physics, Kyoto University, Kyoto 606-8502, Japan}

\author{Bo-Wen Xiao}
\affiliation{Key Laboratory of Quark and Lepton Physics (MOE) and Institute
of Particle Physics, Central China Normal University, Wuhan 430079, China}

\author{Feng Yuan}
\affiliation{Nuclear Science Division, Lawrence Berkeley National
Laboratory, Berkeley, CA 94720, USA}

\begin{abstract}
We investigate the contributions from the so-called elliptic gluon Wigner distributions to
the rapidity and azimuthal correlations of particles produced in high energy $pp$ and $pA$ collisions by 
applying the double parton scattering mechanism. We compute the `elliptic flow' parameter $v_2$ as a function of the transverse momentum and rapidity, and find qualitative agreement with experimental observations. 
This shall encourage 
further developments with more rigorous studies of the elliptic gluon distributions and  
their applications in hard scattering processes in $pp$ and $pA$ collisions. 
\end{abstract}
\maketitle

\section{Introduction}

One of the interesting experimental observations from the proton-proton and proton-nucleus
collisions at the Large Hadron Collider (LHC) and Relativistic Heavy Ion Collider (RHIC) is 
the long range rapidity and azimuthal angle correlations between 
hadrons~\cite{Khachatryan:2010gv,Abelev:2012ola,Aad:2012gla,CMS:2012qk,Adare:2014keg,
Adare:2015ctn,Aad:2014lta,Aad:2015gqa,Aaboud:2016yar}, see, e.g., a recent review in Ref.~\cite{Dusling:2015gta}.
These intriguing observations have generated great theoretical investigations, and many models
have been proposed to explain the experimental results, including (but not limited to) 
hydrodynamics~\cite{dEnterria:2010xip,CasalderreySolana:2009uk,Avsar:2010rf,Schenke:2014zha,Yan:2013laa}, 
QCD motivated models~\cite{Kopeliovich:2007fv,Kopeliovich:2008nx,Gyulassy:2014cfa,Gambini:2016ijt}, and in particular,  the multi-gluon
correlations calculated in the Color Glass Condensate (CGC) framework~\cite{Kovner:2010xk,
Kovner:2011pe,Dusling:2012wy,Dusling:2012cg,
Kovchegov:2012nd,Iancu:2013uva,Schenke:2015aqa,Dumitru:2014dra,
Dumitru:2010mv,Dumitru:2014yza,Lappi:2015vha,Levin:2011fb,Gotsman:2016whc,Lappi:2015vta}.

In this paper, we investigate the contribution from the double parton scattering (DPS)~\cite{Diehl:2011yj,Manohar:2012pe}
coupled with the so-called elliptic gluon Wigner distribution~\cite{Hatta:2016dxp,Hagiwara:2016kam,Zhou:2016rnt}. 
In high energy collisions, we expect the DPS, or in general, the multi-parton scattering, is
the dominant source for multi-particle productions. A unique feature of DPS is that its contribution
is not strongly suppressed for near-side particle productions with large rapidity separation 
as compared to the single parton scattering (SPS) contribution. Therefore, DPS
may well be the dominant source for long range correlations among produced hadrons. 

It was first pointed out in Ref.~\cite{Strikman:2010bg} that the DPS plays an important role in two 
particle production in forward $pA$ and $dA$ collisions
at RHIC. This idea was followed up in the saturation formalism in Ref.~\cite{Stasto:2011ru} to estimate
the so-called pedestal contribution in the correlation measurements. Further study 
in Ref.~\cite{Lappi:2012nh} also confirmed the importance of these contributions in the two 
particle production in $pA$ collisions. However, all these studies assumed that the
two hard scatterings are essentially uncorrelated. In the following, we will extend the DPS
mechanism to include the impact parameter dependence which naturally encodes the
correlation between the two scatterings. If we average over the impact parameter
space, this will reduce to the previous applications of the DPS mechanism in the CGC
framework. However, the unintegrated gluon distribution involved in these scatterings
depends on the impact parameter. In particular, there is a nonzero $\cos(2\phi)$ azimuthal correlation 
between the transverse momentum $k_\perp$ and the impact parameter $b_\perp$,
which was referred to as the elliptic gluon Wigner distribution in Ref.~\cite{Hatta:2016dxp}.
Since the impact parameters for the two hard scatterings are correlated due to the DPS 
mechanism, we expect the transverse momenta from the two hard scatterings are
correlated as well. This will naturally give rise to the $\cos(2\phi)$ two-particle correlation in the final state.

In Ref.~\cite{Hatta:2016dxp}, the elliptic gluon Wigner distribution has been shown to be measurable in
diffractive dijet production in lepton-nucleon collisions at  the future
electron-ion collider (EIC). The present study suggests that the same distribution can affect various observables in different types of collisions.   


The rest of this paper is organized as follows. In Sec.~II, we study the DPS
contributions to the two particle production in the dilute-dense collisions and derive a formula for the `elliptic flow' parameter $v_2$.
The result is relevant to $pp$ and $pA$ experiments at RHIC and the LHC. 
 In Sec.~III, we numerically evaluate $v_2$ in a model which incorporates the saturation effect in the target.
We point out some generic features of the DPS contributions which can be
compared to the experimental observations.
We summarize our paper in Sec.~IV.

\section{Double Parton Scattering Contributions in the Dilute-Dense Collisions}


In order to describe the near-side two particle correlations in $pp$ and $pA$ collisions, we  
 introduce the impact parameter dependence in the DPS framework. 
Similarly to the derivation of DPS in Refs.~\cite{Diehl:2011yj,Manohar:2012pe},
we write down the generic expression for the differential cross 
section of two parton production as
\begin{eqnarray}
&&  \left. \frac{d\sigma}{dy_1d^2k_{1\perp}dy_2d^2k_{2\perp}}
\right|_{DPS}\nonumber \\
&&=\int d^2x_\perp d^2y_\perp d^2b_{1\perp} d^2b_{2\perp} e^{ik_{1\perp}\cdot x_\perp}e^{ik_{2\perp}\cdot y_\perp}F_A(x_p,x_p';z_\perp)F_B(x_A,x_A';\vec{b}_{1\perp},\vec{b}_{2\perp};\vec{x}_\perp,\vec{y}_\perp) \ ,
\label{1}
\end{eqnarray}
where $z_\perp=|\vec{b}_{1\perp}-\vec{b}_{2\perp}|$, and $\vec{b}_{1\perp}$ and $\vec{b}_{2\perp}$
denote the two hard scattering positions with respect to the center of the target.  The  `dipole sizes' $x_\perp$ and $y_\perp$ are 
Fourier-conjugate variables to the partons' outgoing transverse momentum $k_{1\perp}$ and $k_{2\perp}$, respectively.  The longitudinal momentum fractions $x_p$, $x_p'$, $x_A$, and $x_A'$ are determined by the final state kinematics. 
The physics picture is that two partons from the incoming proton encounter 
multiple scattering off the target, and fragment into two final state particles. 
The multiple scattering  is described in the CGC framework or in the color-dipole 
model. For a large nucleus, 
we can assume a factorized form 
\begin{eqnarray}
F_B\approx S_{x_A}\left(\vec{b}_{1\perp},\vec{x}_\perp\right) S_{x_A'}\left(\vec{b}_{2\perp},\vec{y}_\perp\right)\,, \label{fact}
\end{eqnarray}
where $S$ is the dipole S-matrix which may be in the fundamental or adjoint representation  depending on the
partonic channels involved in the DPS. The terms neglected in (\ref{fact}) are of order $1/N_c^2$. It has been argued  \cite{Lappi:2015vta} that these color-suppressed, but `connected' contributions can give rise to nonvanishing $v_2$ in $pp$ and $pA$ collisions. Moreover, if the the target is small, as in $pp$ collisions, factorization (\ref{fact}) is violated even in the large-$N_c$ limit due to the small-$x$ evolution in the target. (In the case of a dipole target, this can be shown analytically \cite{Hatta:2007fg,Avsar:2008ph}.) Such factorization breaking effects have been considered as another source of $v_2$ in small systems \cite{Kovner:2010xk,Kovner:2011pe,Kovner:2012jm}.

Here we show that, even if the factorization (\ref{fact}) holds strictly, there exist non-trivial angular correlations between the two 
outgoing particles due to the angular correlation between $\vec{b}_{1_\perp}$ and $\vec{x}_\perp$ in the $S$-matrix. It should be mentioned that the idea that the correlation between impact parameter and dipole orientation generates  anisotropy in the final state  has been previously studied in the context of single \cite{Kopeliovich:2007fv,Kopeliovich:2008nx}  (see also, \cite{Iancu:2017fzn})  and double \cite{Levin:2011fb,Gotsman:2016whc} parton scattering. Thus, the approach here is essentially the same  as in \cite{Levin:2011fb,Gotsman:2016whc}.  Yet, our formulation is considerably more concise and clearly establishes the connection to the elliptic gluon Wigner distribution which is a fundamental object in the tomographic study of the nucleon/nucleus.   
 
For this purpose, let us write (\ref{1}) as 
\begin{eqnarray}
 \left. \frac{d\sigma}{dy_1d^2k_{1\perp}dy_2d^2k_{2\perp}}
\right|_{DPS}=\int  d^2b_{1\perp} d^2b_{2\perp} F_A(x_p,x_p';z_\perp)G_{x_A}(\vec{b}_{1\perp},\vec{k}_{1\perp}) G_{x_A'}(\vec{b}_{2\perp},\vec{k}_{2\perp}) \ , \label{mod}
\end{eqnarray}
 where $G(\vec{b}_\perp,\vec{k}_\perp)$ is the Fourier transform of $S(\vec{b}_\perp,\vec{x}_\perp)$ and we assumed (\ref{fact}). The angular correlation between $\vec{b}_\perp$ and $\vec{x}_\perp$ is transformed into the one between $\vec{b}_\perp$ and $\vec{k}_\perp$. At small-$x$, this correlation is dominantly {\it elliptic} \cite{Hatta:2016dxp,Hagiwara:2016kam}, namely, 
\beq
G(\vec{b}_\perp,\vec{k}_\perp) = G^0(b_\perp,k_\perp) + 2\cos 2(\phi_b-\phi_k) \widetilde{G}(b_\perp,k_\perp) + \cdots \,. \label{el}
\eeq 
The angular integrals in (\ref{mod}) then lead to an elliptic angular correlation of the form $\cos 2(\phi_{k_1} -\phi_{k_2})$. 

This can be seen most clearly and model-independently at large impact parameter where it is convenient to  write 
 $\vec{b}_{1,2\perp} = \vec{b}_\perp \pm \vec{z}_\perp/2$,  
 so that $d^2b_{1\perp}d^2b_{2\perp}=d^2z_\perp d^2b_\perp$. Since the two partons are confined in the proton, the $z_\perp$ integral is limited within the confinement radius $z_\perp\lesssim 1/\Lambda$. When $b_\perp \gg 1/\Lambda\sim z_\perp$, we can  approximately integrate over $z_\perp$ 
to obtain the collinear double parton distribution of the proton,
\begin{equation}
\int d^2z_\perp F_A(x_p,x_p';z_\perp)={\cal D}_{p}(x_p,x_p')\ , 
\end{equation}
which can be further simplified as 
${\cal D}_{p}(x_p,x_p') ={\cal C}(x_p,x_p')f(x_p) f(x_p')$
with ${\cal C}\approx 1$.
With this approximation, we can write down the differential cross section
as 
\begin{eqnarray}
 &&\left. \frac{d\sigma}{dy_1d^2k_{1\perp}dy_2d^2k_{2\perp}}
\right|_{DPS}\sim \int_{1/\Lambda} d^2b_\perp f(x_p)f(x_p') G_{x_A}(\vec{b}_\perp,\vec{k}_{1\perp}) G_{x_A'}(\vec{b}_{\perp},\vec{k}_{2\perp}) \\ 
&& \propto  \pi \int_{1/\Lambda}\!\! db_\perp^2  \left[G^0_{x_A}(b_\perp,k_{1\perp})G^0_{x_A'} (b_\perp, k_{2\perp}) 
+2\cos2(\phi_{k_{1\perp}}\!-\!\phi_{k_{2\perp}})\widetilde{G}_{x_A}(b_\perp, k_{1\perp})\widetilde{G}_{x_A'} (b_\perp,k_{2\perp})\right]  \,.  \nonumber
\end{eqnarray}
As expected, we recognize the $\cos 2(\phi_{k_1}-\phi_{k_2})$ correlation proportional to the elliptic part $\widetilde{G}$ squared. 

We now turn to the small impact parameter region $b_\perp\sim z_\perp \sim 1/\Lambda$. To proceed, we introduce a Gaussian model $F_A(z_\perp) \propto e^{-z_\perp^2 \Lambda^2}$. 
The angular integrals can then be performed as
\beq
&& \int^{1/\Lambda} d^2b_{1\perp} d^2b_{2\perp} e^{-\Lambda^2 |\vec{b}_{1\perp}-\vec{b}_{2\perp}|^2} G_{x_A}(b_{1\perp},k_{1\perp})G_{x_A'}(b_{2\perp},k_{2\perp})  \nonumber \\
&&=4\pi^2 \int_0^{1/\Lambda} b_{1\perp}db_{1\perp} b_{2\perp}db_{2\perp} e^{-\Lambda^2 (b_{1\perp}^2 + b_{2\perp}^2)} \Bigl[I_0(2\Lambda^2 b_{1\perp}b_{2\perp})G^{0}_{x_A}(b_{1\perp},k_{1\perp})G^0_{x_A'} (b_{2\perp},k_{2\perp}) \nonumber \\  
&& \qquad \qquad  +2\cos2(\phi_{k_{1\perp}}-\phi_{k_{2\perp}})I_2(2\Lambda^2 b_{1\perp}b_{2\perp}) \widetilde{G}_{x_A}(b_{1\perp},k_{1\perp})\widetilde{G}_{x_A'} (b_{2\perp,}k_{2\perp})\Bigr]\,. \label{last}
\eeq
We again find the elliptic correlation $\cos 2(\phi_{k_1}-\phi_{k_2})$. Other models of $F_A$ will also give rise to this correlation, as long as $F_A$ depends on the angle between $\vec{b}_{1\perp}$ and $\vec{b}_{2\perp}$ via $z_\perp = |\vec{b}_{1\perp}-\vec{b}_{2\perp}|$.

Noting that the upper limit of the  $b_{1,2\perp}$-integrations in (\ref{last}) can actually be extended to some value $R_{cut}>1/\Lambda$, we define 
\beq
V_2(k_{1\perp},k_{2\perp}) \equiv \frac{\int_0^{R_{cut}} b_{1\perp}db_{1\perp} b_{2\perp}db_{2\perp} e^{-\Lambda^2 (b_{1\perp}^2 + b_{2\perp}^2)} I_2(2\Lambda^2 b_{1\perp}b_{2\perp}) \widetilde{G}_{x_A}(b_{1\perp},k_{1\perp})\widetilde{G}_{x_A'} (b_{2\perp},k_{2\perp}) }{ \int_0^{R_{cut}} b_{1\perp}db_{1\perp} b_{2\perp}db_{2\perp} e^{-\Lambda^2 (b_{1\perp}^2 + b_{2\perp}^2)} I_0(2\Lambda^2 b_{1\perp}b_{2\perp})G^0_{x_A}(b_{1\perp},k_{1\perp})G^0_{x_A'} (b_{2\perp},k_{2\perp}) }\,,\nonumber \\ 
\label{v}
 \eeq
This is related to the experimentally measured $v_2$ via  
\beq
v_2(k_\perp,k_\perp^{ref}) \equiv \frac{V_2(k_\perp,k_\perp^{ref})}{\sqrt{V_2(k_\perp^{ref},k_\perp^{ref})}}\,, \label{fac}
\eeq
where $k_\perp^{ref}$ denotes some reference momentum. 
Experimentally, it has been observed that $v_2$ is roughly independent of $k_\perp^{ref}$. This implies that the two-particle correlation $V_2$ factorizes $V_2(k_{1\perp},k_{2\perp}) \approx v_2(k_{1\perp})v_2(k_{2\perp})$, and this is consistent with the hydrodynamic interpretation of $v_2$. In our case (\ref{v}), the integrand approximately factorizes at small $b_{1,2\perp}$. However, the result after the full  $b_{1,2\perp}$-integrations does not factorize in general.

\section{Model calculation}

To illustrate the DPS contribution discussed above, we evaluate (\ref{v}) in a model that incorporates the gluon saturation effect at small-$x$. For definiteness, we consider the $pp$ collisions partly because realistic models for $pA$ with both the $b_\perp$-dependence and the small-$x$ evolution are not available to us.   
The angular independent part $G^0$ and the elliptic part $\widetilde{G}$ are computed from the solution of the impact parameter dependent Balitsky-Kovchegov (BK) equation in the same way as explained in \cite{Hagiwara:2016kam}. The only difference is that here we use a different initial condition to be slightly more realistic ($e^{-d^2} \to e^{-cd^2}$ with $c= 6$ in Eq.~(7) of \cite{Hagiwara:2016kam}). For small dipole sizes $r_\perp\to 0$ and at small impact parameter $b_\perp \approx 0$, this gives the initial condition   $S_{Y=0}(b_\perp,r_\perp) \approx e^{-cr_\perp^2/R^2}$ where $R$ is roughly the size of the target. For the proton, we use $R=1\,$fm so that the initial condition is $\sim e^{-r_\perp^2/(0.4{\rm fm})^2}$ which is the same as the original GBW model \cite{GolecBiernat:1998js}.    
Below we use $R$ as the unit of length, so for example $R_{cut}=2$ means $R_{cut}=2R$. The current numerical result is intended to be viewed as an illustrative example which shows that this mechanism due to the elliptic Wigner distribution can generate sizeable elliptic flow in small systems. Of course, a more realistic numerical model calculation should be carried out in the future in order to compare with the experimental data measured for small systems created in $pp$ and $pA$ collisions. 

The results for $k_\perp=k_\perp^{ref}$ and $x_A=x_{A'} = e^{-Y}$ are shown in Fig.~1 for different values of $Y$ with $\Lambda=1$. In the two figures, we used different values of the cutoff $R_{cut}$ in the $b_\perp$-integral.  Actually, the elliptic part $\widetilde{G}$ becomes negative in the large-$k_\perp$ region. This means that the squared function $(\tilde{G})^2$ in the numerator has two peaks in the $k_\perp$ direction at fixed $b_{1\perp}\approx b_{2\perp}$. However, the zero of $\tilde{G}$ is $b_\perp$-dependent,  so that after integrating over $b_\perp$ we obtain the single-peak structure as in the right figure. On the other hand, if the cutoff is small as in the left figure, $b_\perp$-averaging is insufficient and we see a remnant of the double-peak structure. Note that the peak position of $v_2$ is almost independent of $Y$, and does {\it not} coincide with the saturation momentum $k\sim Q_s(Y)$ which is a rapidly increasing function of $Y$. This is a characteristic property of the elliptic part $\widetilde{G}$ observed in  
 \cite{Hagiwara:2016kam}.  Interestingly, Ref.~\cite{Kovner:2012jm} argued that the peak position of $v_2$ from the present mechanism  should occur at the inverse correlation length of $Q_s$ in the transverse plane which is much smaller than $Q_s$. It remains to be seen whether the observation in  \cite{Hagiwara:2016kam} can be physically interpreted as such. 
We also note that the height of the peak decreases with increasing $Y$ largely because  $\widetilde{G}$ has the same property.  
Compared with the experimental data, the magnitude of $v_2$ in our model is somewhat smaller even for the smallest values of $Y$.

Next we test the degree of factorization. We choose $k_\perp \neq k_{\perp}^{ref}$ and check the $k_{\perp}^{ref}$-dependence of the result. The integrand of (\ref{v}) factorizes at small $b_\perp$, because $I_2(2\Lambda^2 b_{1\perp}b_{2\perp}) \sim (b_{1\perp}b_{2\perp})^2$. Thus, factorization is good if $R_{cut}$ is small and this is clearly seen in Fig.~\ref{fig2} ($Y=4$) and Fig~\ref{fig3} ($Y=8$). 
We also see that the factorization holds better for larger $Y$ values. Note that $v_2$ is no longer positive definite once we allow $k_\perp \neq k_\perp^{ref}$ because $\widetilde{G}$ is negative for large $k_\perp^{ref}$. When $Y$ becomes large, the negative region of  $\widetilde{G}$ is pushed to a larger $k$-region and $v_2$ tends to become positive.

Finally we study the rapidity correlation between two particles. We take $x_A$ and $x_{A'}$ in (\ref{v}) to be different  and compute $v_2(k_\perp=k_\perp^{ref}=2)$ from (\ref{fac}) as a function of $\Delta Y=Y-Y^{ref} = \ln \frac{1}{x}-\ln \frac{1}{x^{ref}}$. (In practice we set $Y^{ref}=0$.) The result is shown in Fig.~\ref{fig4}. We recognize a certain degree of `long-range rapidity correlation', namely, $v_2$ decreases only slowly with increasing $\Delta Y$. This is largely due to approximate factorization of the double scattering amplitude (\ref{fact}).  We should mention that when $\Delta Y$ becomes too large such that $\alpha_s \Delta Y\sim 1$, one has to consider the small-$x$ evolution between the two rapidities \cite{Iancu:2013uva}. This is however beyond the scope of this work.

\begin{figure}[tbp]
\begin{center}
\includegraphics[width=8cm]{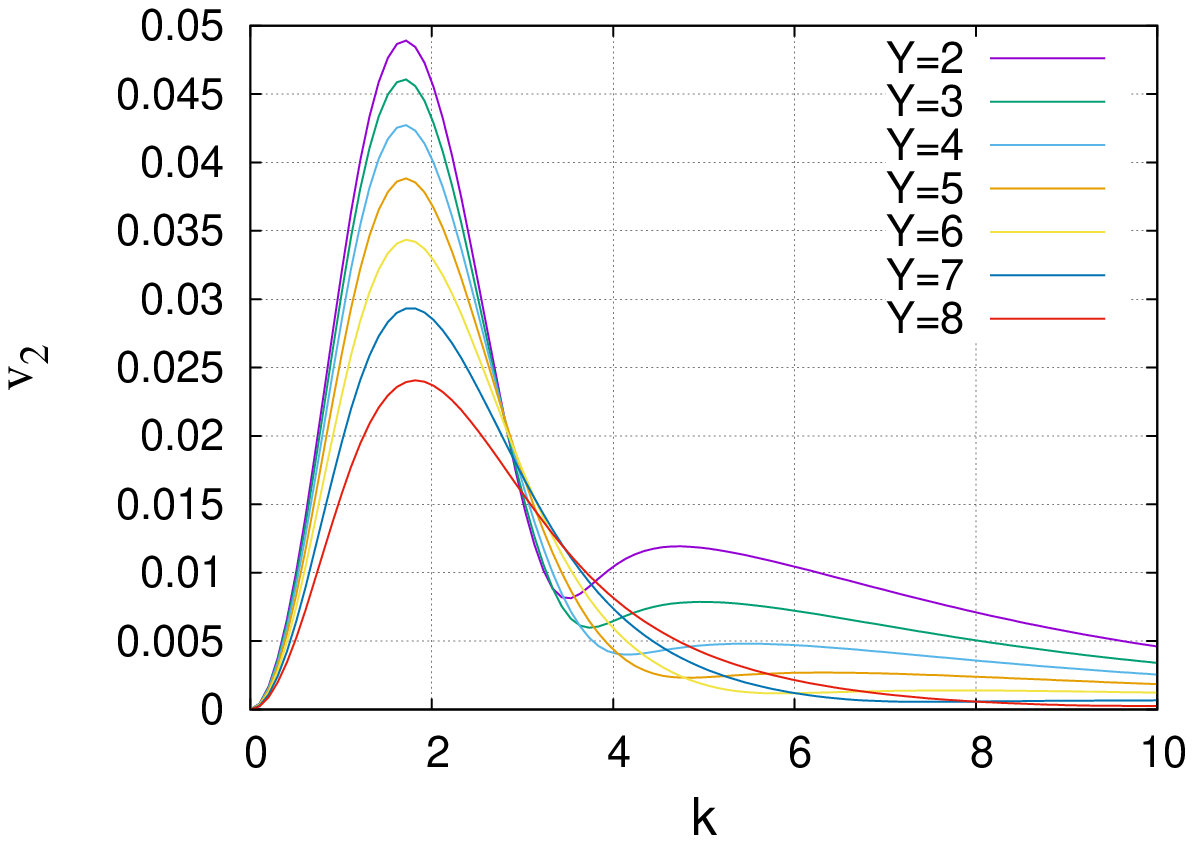}
\includegraphics[width=8cm]{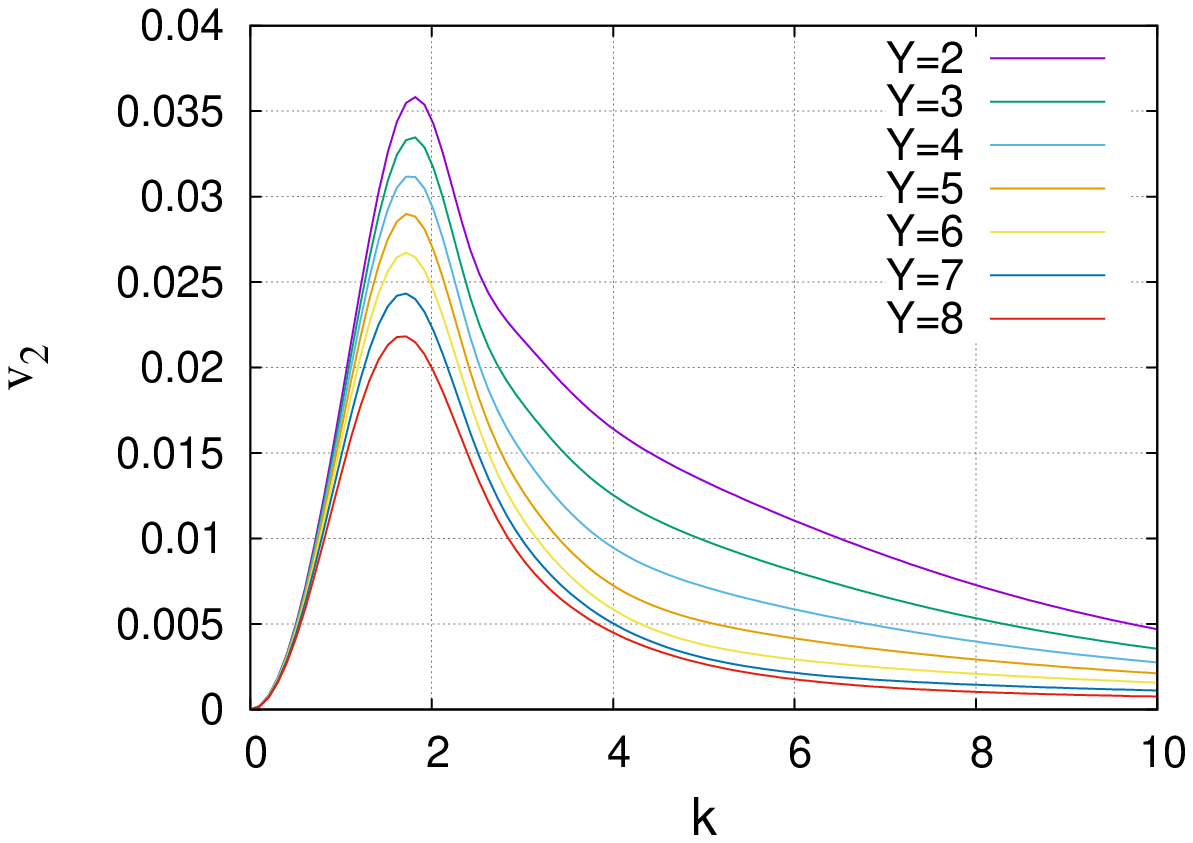}
\end{center}
\vspace{3mm}
\caption[*]{$v_2$ as a function of $k_\perp$. The cutoff in the $b_\perp$-integration in (\ref{v}) is $R_{cut}=2$ (left) and $R_{cut}=5$ (right).
}
\label{fig1}
\end{figure}

\begin{figure}[tbp]
\begin{center}
\includegraphics[width=8cm]{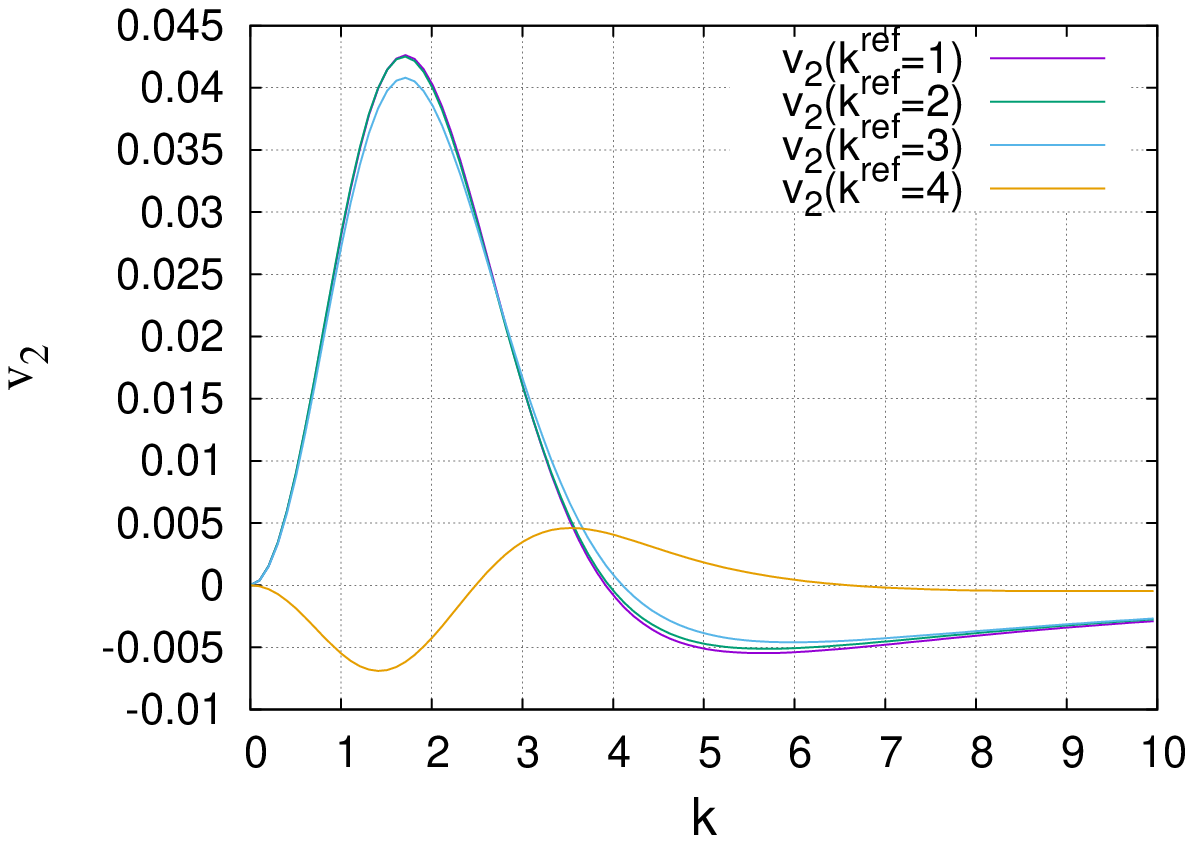}
\includegraphics[width=8cm]{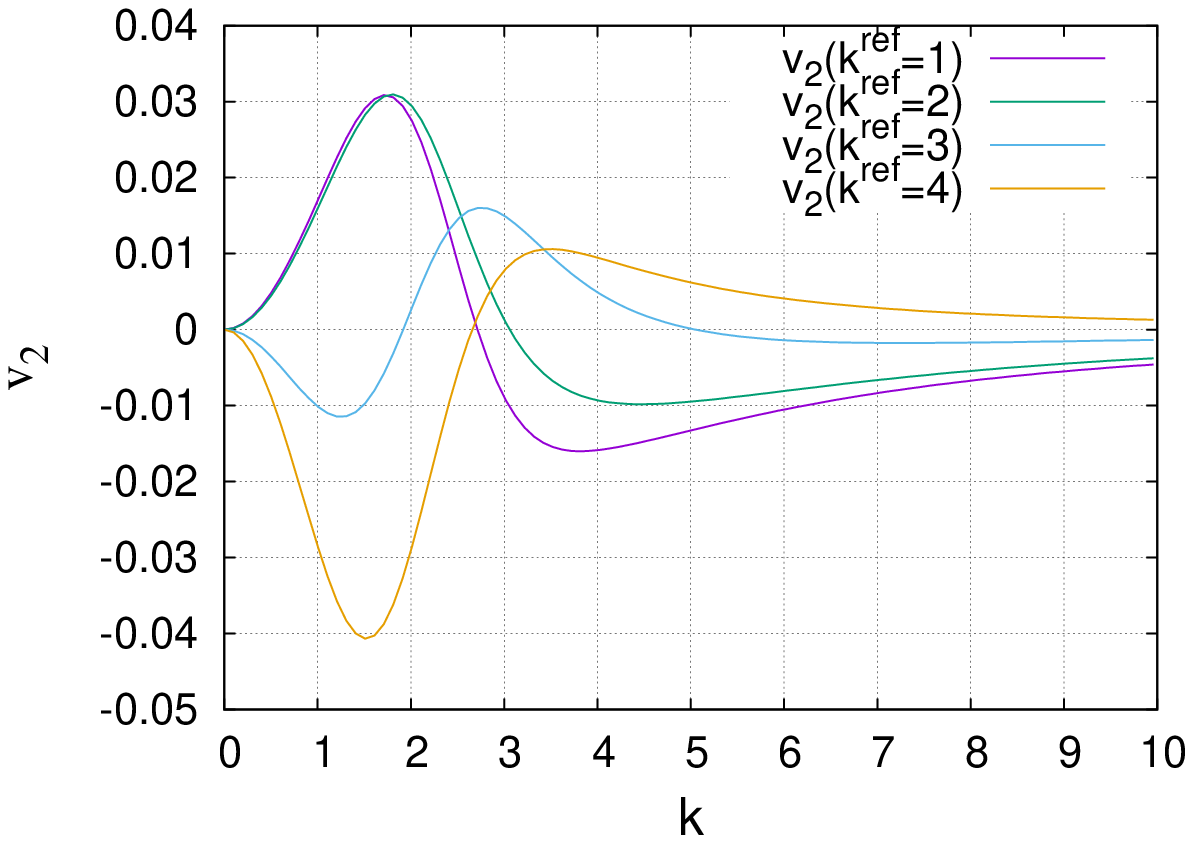}
\end{center}
\vspace{3mm}
\caption[*]{$k^{ref}_\perp$-dependence at $Y=4$. Left: $R_{cut}=2$, Right: $R_{cut}=5$. }
\label{fig2}
\end{figure}

\begin{figure}[tbp]
\begin{center}
\includegraphics[width=8cm]{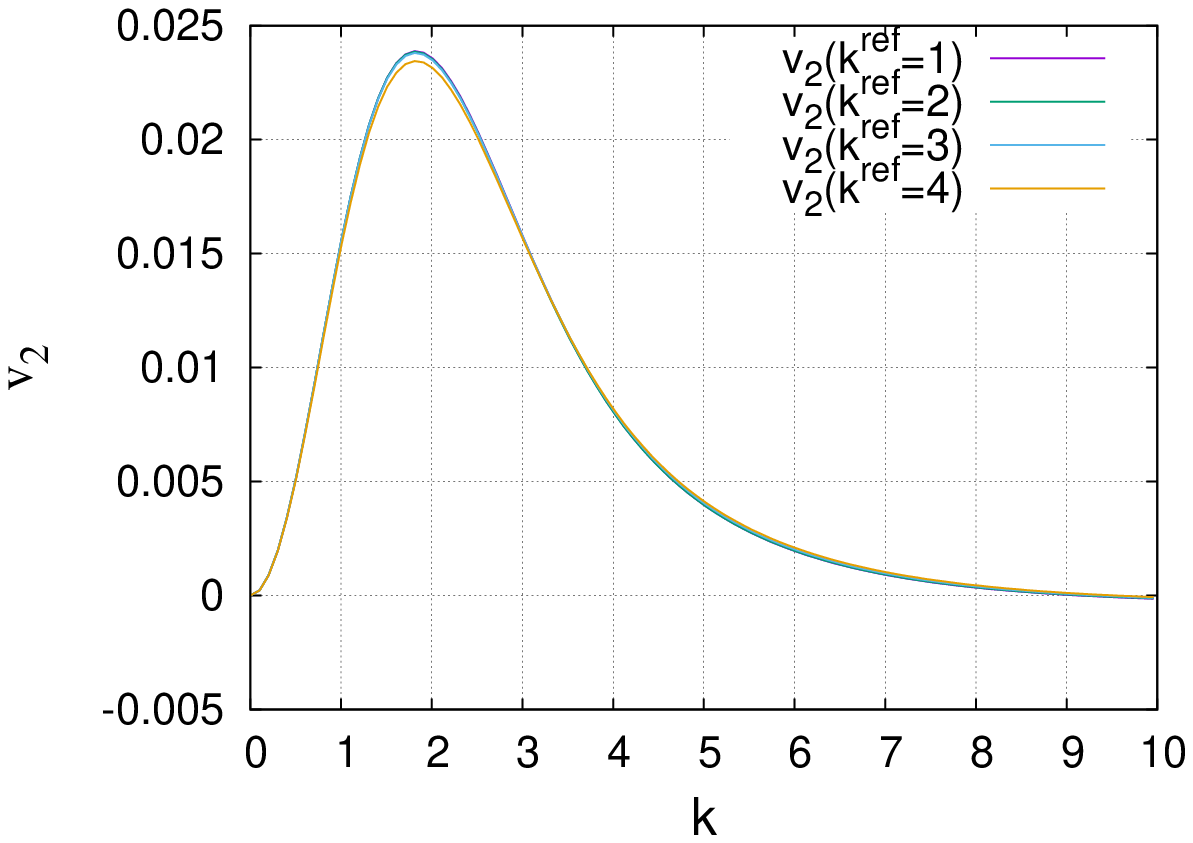}
\includegraphics[width=8cm]{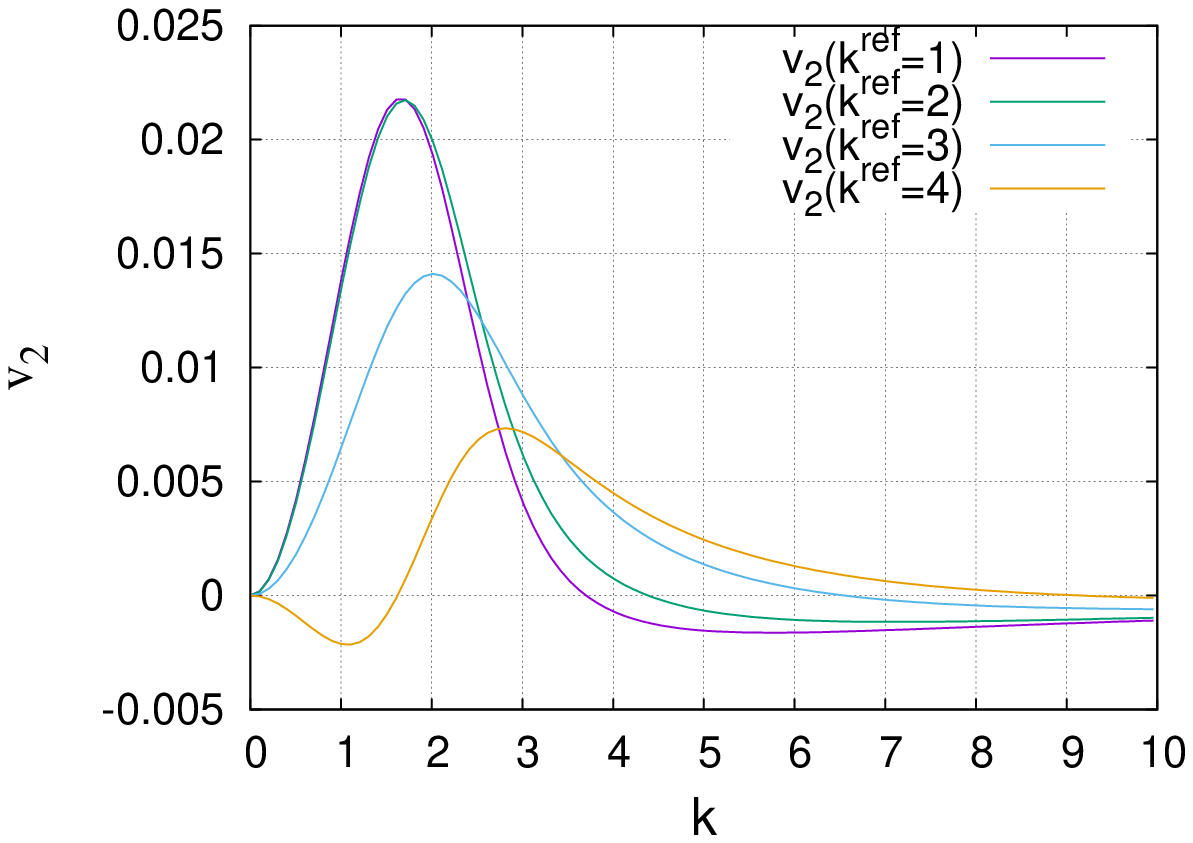}
\end{center}
\vspace{3mm}
\caption[*]{$k^{ref}_\perp$-dependence at $Y=8$. Left: $R_{cut}=2$, Right: $R_{cut}=5$. }
\label{fig3}
\end{figure}


\begin{figure}[tbp]
\begin{center}
\includegraphics[width=8cm]{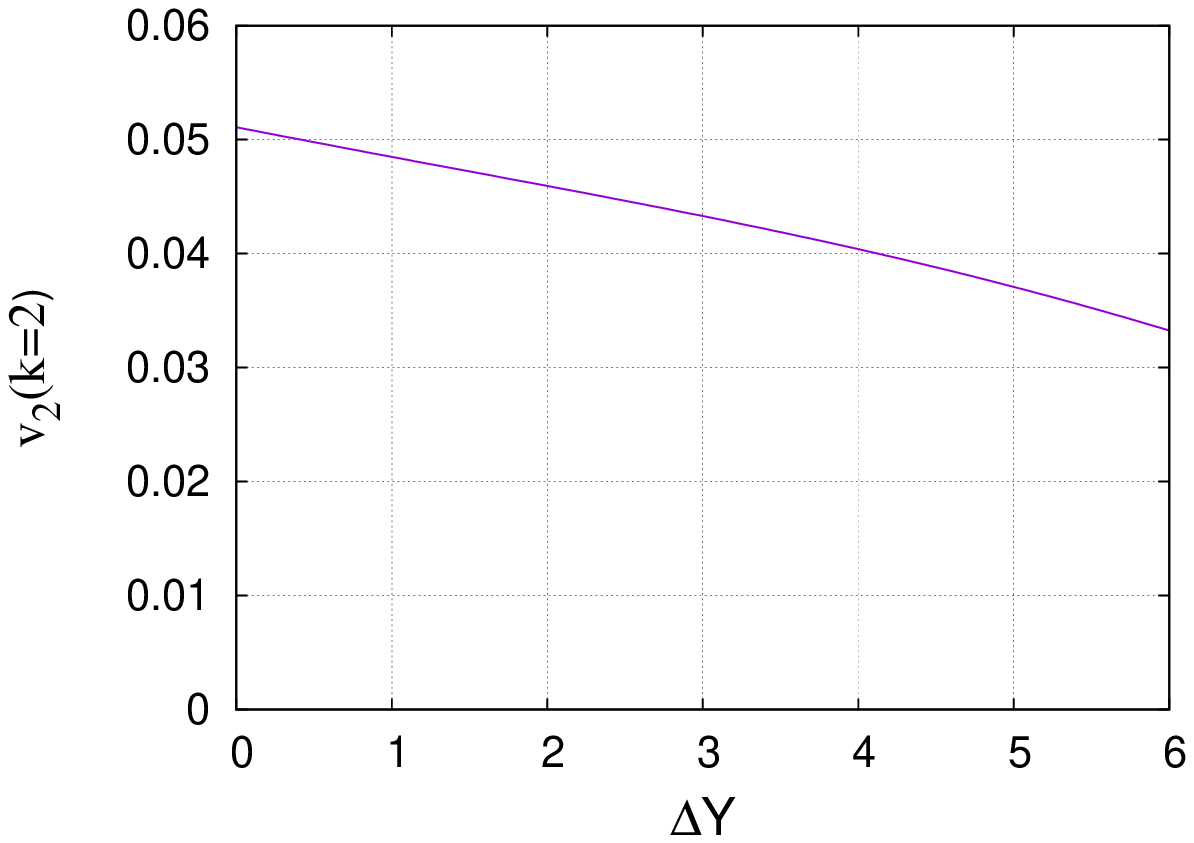}
\includegraphics[width=8cm]{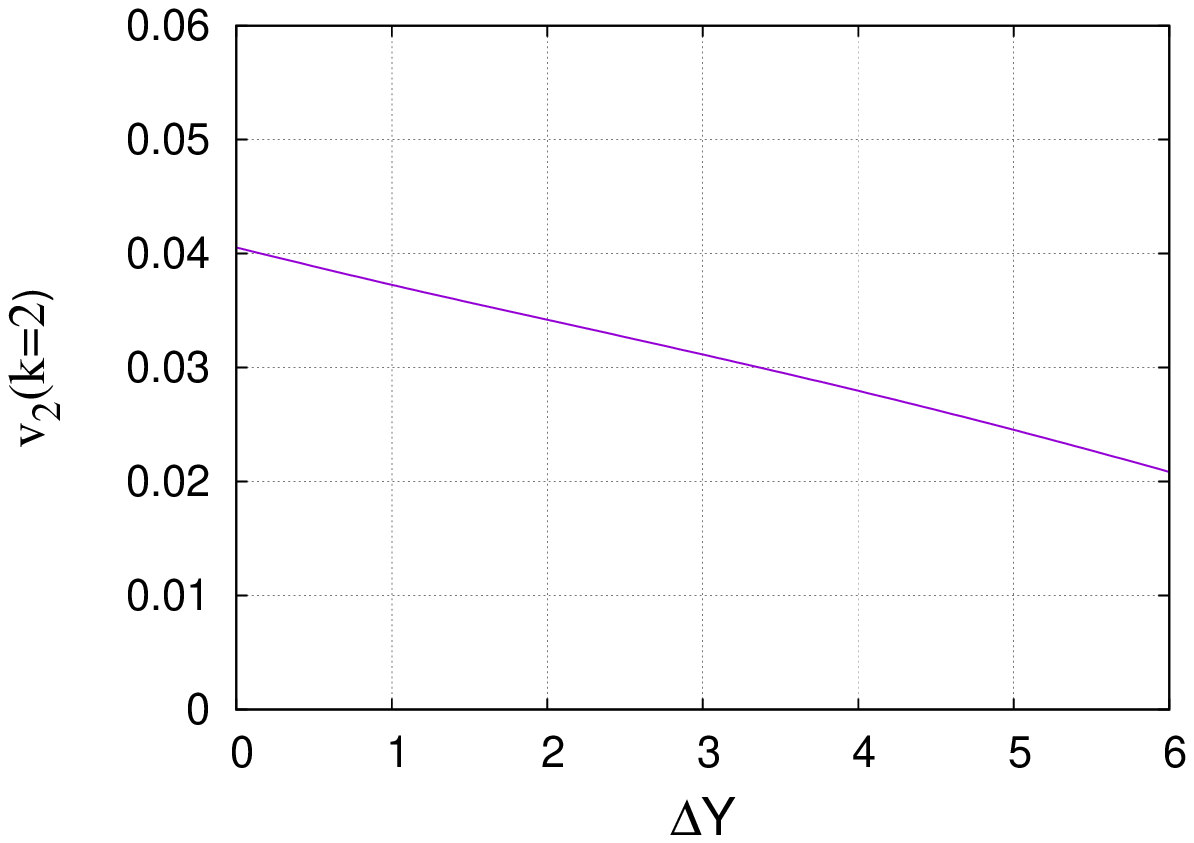}
\end{center}
\vspace{3mm}
\caption[*]{$v_2(k_\perp=k_\perp^{ref}=2)$ as a function of the rapidity difference $\Delta Y$ between two particles.  Left: $R_{cut}=2$, Right: $R_{cut}=5$. }
\label{fig4}
\end{figure}

\section{Conclusions and Discussions}

To summarize, we have explored the two-particle productions in small systems created in high energy $pp$ and $pA$
collisions from the double parton scattering mechanism, where the elliptic gluon Wigner 
distributions give rise to the desired $\cos(2\phi)$ azimuthal angular correlations. 
By applying the DPS idea, we derived a formula for the two particle correlation in the small-$x$
saturation formalism, where two partons from the incoming nucleon scatter with
the target nucleon/nucleus and produce two particles in the final state. The DPS mechanism imposes
the impact parameter correlation between the two hard partonic scattering 
processes, which results in a correlation between the transverse momenta 
of the final state particles. Due to the unique feature of the DPS mechanism, this
correlation will not decrease dramatically with the increase of the rapidity 
difference between the two particles. 

We have also applied a recent result of the elliptic gluon distributions from the
BK equation~\cite{Hagiwara:2016kam}
to illustrate their contributions to the $\cos(2\phi)$ correlation between two particles produced 
in high energy $pp$ and $pA$ collisions. The size of the elliptic flow parameter $v_2$
was found in a similar range as the experimental observations at RHIC 
and the LHC. This is an encouraging message, and demonstrates that the
long range correlation of $v_2$  may have significant contributions 
from the elliptic gluon distributions in the target. However, we would like to emphasize that we 
have limited knowledge on the elliptic gluon distribution as compared to the usual 
dipole gluon distribution, and more studies are needed to compare to the experimental 
data. This will also help to pin down the underlying mechanism for the novel ``flow" 
phenomena in high energy $pp$ and $pA$ collisions. 

Finally, we would like to point out that the elliptic gluon Wigner distribution represents
a nontrivial tomography structure of gluons inside the nucleon/nucleus. Its dependence
on $x$, $b_\perp$ and $k_\perp$ will provide not only the imaging of parton distributions
at small-$x$, but also the unique opportunity to explore the QCD dynamics 
associated with the small-$x$ evolution~\cite{Hagiwara:2016kam}. 
Since the elliptic gluon distribution can be well studied in  hard
diffractive dijet production at the EIC~\cite{Hatta:2016dxp}, the comparison between the two 
particle elliptic flow in $pp$ and $pA$ collisions and further observables in $eA$ collisions will be crucial
to understand the gluon dynamics under extreme conditions. We hope to come back to this
issue soon.

\acknowledgments
We thank Edmond Iancu and Amir Rezaeian for correspondence about their related work \cite{Iancu:2017fzn}.
This material is based upon work supported by the U.S. Department of Energy, 
Office of Science, Office of Nuclear Physics, under contract number 
DE-AC02-05CH11231 and by the Natural Science Foundation of China (NSFC) under Grant No.~11575070.


\end{document}